\documentclass[aps,prx,twocolumn,amsmath,amssymb,floatfix,superscriptaddress]{revtex4-2} %rmp

\usepackage{amsmath,amssymb,amsfonts,graphicx,bm,array,physics}
\usepackage{natbib} 
\usepackage{bbm}
\usepackage[outline]{contour}
\usepackage{color}
\usepackage{dcolumn}
\usepackage[utf8]{inputenc}
\usepackage{enumitem}
\usepackage{epstopdf}
\usepackage{amsthm}
\usepackage{empheq}
\usepackage{braket}
\usepackage{mathtools}
\usepackage[usenames,dvipsnames]{xcolor}
\usepackage{latexsym}
\usepackage{hyperref} % citecolor=Cerulean,linkcolor=RubineRed,urlcolor=Cerulean
\hypersetup{colorlinks = true,allcolors = blue }
\usepackage[capitalize]{cleveref}
\usepackage{diagbox} %longtable
\usepackage[caption=false]{subfig}
\usepackage{multirow}
\usepackage{outlines}

\usepackage{epsf}

\usepackage{xspace}

\newcommand{\rob}{\mathcal{R}}
\newcommand{\free}{\mathcal{F}}

\usepackage{comment}

\def\Sep{ \textnormal{Sep} }

\def\Incoh{ \textnormal{Incoh} }
\def\Coh{ \textnormal{Coh} }

\def\Ent{ \textnormal{Ent} }

\renewcommand{\tr}[1]{\operatorname{\textnormal{Tr}}\left( {#1} \right)}

\newcommand{\bseq}{\begin{subequations}}
\newcommand{\eseq}{\end{subequations}}
\newcommand{\bsplit}{\begin{split}}
\newcommand{\esplit}{\end{split}}

\newcommand{\reals}{\mathbb{R}}

\theoremstyle{definition}

\newtheorem{lemma}{Lemma}
\newtheorem{theorem}{Theorem}

\def\Var{\mathrm{Var}}
\def\Cov{\mathrm{Cov}}
\def\Re{\mathrm{Re}}

\newcommand{\logder}{\mathcal{L}}

\begin{document}

\title{%On the certification of quantum resources \\
Witnessing Magic (and other Quantum Resources)
}

%\author{Brian Barch}
%\affiliation{Quantum and Condensed Matter Physics Group (T-4), Theoretical Division, Los Alamos National Laboratory, Los Alamos, New Mexico 87545, USA}

%\author{Sarah}

%\author{xxx}

%\author{xxx}

%\author{Luis Pedro Garc\'ia-Pintos}
%\affiliation{Quantum and Condensed Matter Physics Group (T-4), Theoretical Division, Los Alamos National Laboratory, Los Alamos, New Mexico 87545, USA}
\author{Brian Barch}
\author{Luis Pedro Garc\'ia-Pintos}
\affiliation{Los Alamos National Laboratory, Los Alamos, New Mexico 87545, USA}
\date{\today}

\begin{abstract}
Motivated by measurements of quantum magic, we study the sampling cost of 
resource robustness ---a quantifier for a broad class of quantum resources--- 
by bounding the variance of optimal robustness witnesses. 
We first lower bound the samples required to estimate a state's resource robustness using Cram\'er-Rao bounds. 
We find that the number of measurements scales at least quadratically in the robustness for highly mixed states, and at least linearly in the generalized robustness for broad families of pure states, making it difficult to measure either robustness to low absolute error when the states are highly resourceful.
We then derive two complementary bounds in terms of the geometry of witnesses, one of which is tight for a class of resource theories including coherence and bipartite entanglement. 
An additional hurdle occurs for tracking the onset of resource robustness along state-space trajectories: the integrated witness variance along a path is at least the squared resource gain over Fisher length, indicating that a state evolution with exponential resource gain necessarily goes through states for which the robustness cannot be measured efficiently. 
This hurdle does not occur when estimating robustness to fixed relative, rather than absolute, error. These results suggest that the ideal witness-based resource estimation task, for general resource theories, is the estimation of generalized robustness to low relative error on highly resourceful pure states.
\end{abstract}

\maketitle

\section{Introduction}
 
Quantum computers offer a route to communication, encryption, and materials simulations beyond the reach of classical computation. Understanding our progress towards this goal, from noisy intermediate-scale devices to fault-tolerant architectures, calls for benchmarks that quantify the resources available for quantum information processing. Quantum resource theories provide a unified framework for this task, encompassing magic, entanglement, coherence, non-Gaussianity, non-Markovianity, and other properties of interest~\cite{RobustnessReview,BrandaoPRL2015, anand2019quantifying}. Measuring these resources can inform the assessment of quantum capabilities, and guide the design of the next generation of quantum devices.

Magic is particularly relevant to quantum advantage, as its quantification directly describes the cost of simulating states classically under stabilizer techniques~\cite{howard2017application,bravyi2019simulation,seddon2021quantifying}. Protocols for measuring magic have been demonstrated on quantum processors~\cite{oliviero2022measuring,haug2023scalable,haug2024efficient}, while tensor-network methods enable its calculation in weakly-entangled systems~\cite{tarabunga2024nonstabilizerness}. Hardness results establish substantial barriers to estimating magic in general~\cite{garcia2025hardness,gu2024pseudomagic}, in turn motivating interest in efficiently calculable magic lower bounds~\cite{varela2026predicting, junior2026symmetry, macedo2026littlethingheatdoes}. 

When analyzing the utility of a resource measure, a core question is under what conditions it is accessible within the available computational and experimental resources. Robustness measures are natural candidates because they quantify broad classes of resources and admit a formulation as the maximal expectation value of a resource witness, making them directly measurable. In particular, robustness of magic bounds the non-Clifford resources required for state preparation~\cite{howard2017application}. 
This witness formulation separates two potential challenges to resource estimation: finding a suitable witness, and performing enough measurements to estimate its expectation value to bounded error. Finding or even verifying a magic witness can be computationally difficult~\cite{leone2026unbearable, varela2026predicting}, though it may be easier for other resource theories. Here we focus on the second challenge, and address the sampling cost that remains once an optimal witness is known. Given an optimal witness $W$, the required number of shots to estimate a state's robustness to fixed error $\epsilon$ goes as $\Var(W)/\epsilon^2$, so the shot complexity of the resource robustness is inherently related to the variance of the optimal witness. A high-variance optimal witness can require prohibitively many shots to estimate, making resource estimation inefficient. Witness variance is then a necessary consideration when determining the utility of resource measures towards benchmarking. 

Here, we derive bounds on witness variance for standard and generalized robustness for general quantum resource theories, with magic as principal motivation. We provide background on resource theories, Cram\'er-Rao bounds, and Danskin's theorem in \cref{sec:background}. Building on quantum estimation approaches to entanglement~\cite{genoni2008optimal}, we derive Cram\'er-Rao bounds for mixed states and for the generalized robustness of pure states in \cref{sec:CR-bounds}. In \cref{sec:geometry} we study the geometric nature of pure-state witness variance. Finally, we relate the integrated witness variance along a state-space path to its total robustness gain in \cref{sec:Sampling}, and derive a no-go theorem for efficiently tracking large changes in robustness. We discuss the applicability of the results to benchmarking and directions for future research in \cref{sec:Discussion}.

\section{Background}
\label{sec:background}
\subsection{Resource Theories}

Resource theories are an approach to quantifying the capabilities of quantum states and devices with a broad range of applications~\cite{RobustnessReview}. They divide state space into resourceful states and non-resourceful free states, and, roughly, describe the resourcefulness of a state in terms of its distance from the set of free states.

\subsubsection{Robustness measures}

We consider \emph{robustness measures} to quantify the resources in a quantum system~\cite{RobustnessEntanglement, PhysRevA.67.054305, Cavalcanti_incompatibility,  PhysRevLett.116.150502, RobustnessReview, seddon2021quantifying}. Robustness measures quantify, e.g., resource-enhanced advantages in channel discrimination tasks~\cite{takagiPRL, takagi_general_2019}, and the rates at which resourceful states can be distilled~\cite{BrandaoPRL2015, PhysRevA.101.062315}. Given a convex set $\free$ of free states, the resource robustness is defined by
\begin{align}
\label{eq:Robustness}
\rob( \rho )  &\coloneqq 1+2\cdot\min_
{\gamma \in \free}
\left\{ s \geq 0 \,\, \Big\vert \, \, \frac{\rho + s \gamma }{1+s}  \in \free   \right\}.
\end{align}
Intuitively, $\rob(\rho)$ is a measure of the distance of a given state $\rho$ from $\free$, or, equivalently, it quantifies how much mixing with an arbitrary non-resourceful state $\gamma$ a resource state $\rho$ can withstand while still being resourceful. Sufficient mixing eventually removes all resources from a state. 
It follows that states $\rho$ contained in the free set $\free$ have $\rob(\rho)=1$, while states outside of $\free$ have $\rob(\rho)>1$. 
If the set of free states is convex, the robustness cannot increase under operations that transform free states into free states; i.e., they are monotones under free operations~\cite{RobustnessReview}. For brevity, we will write $\rob$ instead of $\rob(\rho)$ when the state is unambiguous or irrelevant.

The definition of the free set $\free$ characterizes the quantum resource of interest. For instance, if $\free_\Sep$ is defined as the set of separable states, the corresponding $\rob_\Ent$ quantifies a state's entanglement. If $\free_\Incoh$ includes all states without coherence in a given basis, then (a generalization of) $\rob_\Coh$ quantifies coherence in said basis~\cite{PhysRevLett.116.150502}. And most commonly, if $\free_\mathrm{Stab}$ is the set of stabilizer states, then $\rob_\mathrm{Mag}$ quantifies a state's magic.

An operator $W$ is called a resource witness for the free set $\free$ if 
\begin{equation}
    \abs{\tr{W\phi}} \le 1 \quad\forall \phi \in \free.
\end{equation}

The primal definition of robustness admits a dual definition in terms of an optimization problem~\cite{seddon2021quantifying,Regula2018Convex, Winter2022, howard2017application}:
\begin{align}
\label{eq:robustness_dual}
    \rob(\rho) = \max_{W^\dag=W}\left\{\tr{W\rho}\ \Big\vert\ \abs{\tr{W\phi}}\leq1\ \forall\phi\in\free\right\}.
\end{align}
This definition is operationally useful, as it highlights that $\rob$ is measurable, as the expectation value of any optimal witness maximizing the right-hand side of the equation. While the optimal witness set in general depends on state $\rho$, Danskin's theorem (discussed in \cref{sec:bg-danskin}) guarantees that in every direction $x$ in state space, there exists an optimal $W$ such that $\partial_x\braket{W} = \tr{W\partial_x\rho}$ to first order. For this $W$, small changes in the expectation value entirely reflect changes in $\rho$, rather than changes in the witness, so $W$ acts as an estimator of $\rob$ which is unbiased in $x$. One could alternatively probe robustness with any witness, with the resulting expectation value acting as a lower bound on robustness.

A generalization of robustness, called generalized robustness, is defined by restricting the set of witnesses to positive semidefinite operators~\cite{seddon2021quantifying}
\begin{align}
\label{eq:SDP}
    \rob^+(\rho) = \max_{W\geq0}\left\{\tr{W\rho}\ \Big\vert\ \tr{W\phi}\leq1\ \forall\phi\in\free\right\},
\end{align}
in which case the optimal witness can be found via a semidefinite optimization problem~\cite{PhysRevA.72.022310}. 
Since the set of semidefinite witnesses is a subset of the set of Hermitian witnesses, it holds that $\rob^+ \leq \rob$. Additionally, all robustness results derived here also hold for generalized robustness, since they come from properties of the witness set. 
The primal definition of generalized robustness is equivalent to \cref{eq:Robustness}, but $\gamma$ is allowed to be an arbitrary state and the factor of 2 is dropped~\cite{seddon2021quantifying}. In a sense, it quantifies robustness against even resourceful noise. 

\subsubsection{Free-state fidelity}

The stabilizer fidelity measure~\cite{bravyi2019simulation,seddon2021quantifying}, traditionally used in the context of magic,  can be extended to a free-state fidelity  defined for arbitrary resource theories. The free-state fidelity of a pure state $\ket{\psi}$ is then defined as
\begin{align}
\label{eq:free-fidelity}
    F(\psi) \coloneqq \max_{\phi \in \free} \bra{\psi} \phi \ket{\psi}, 
\end{align}
where we use the pure state notation $\psi:=\ketbra{\psi}{\psi}$. The smaller the free-state fidelity $F(\psi)$, the higher the resource in $\ket{\psi}$. 
To see how to relate $F$ to $\rob$, notice that $W_\psi \coloneqq \psi/F(\psi)$ is a feasible, typically sub-optimal, witness for the generalized robustness~\cite{seddon2021quantifying}. Since $\braket{W_\psi}_\psi = 1/F(\psi)$, the hierarchy of resource measures is $1/F \leq \rob^+ \leq \rob$,
with equality of the first two terms when $W_\psi$ is an optimal witness.

\subsection{Cram\'er-Rao Bound}

The Cram\'er-Rao (CR) bound lower-bounds the variance of estimators of parameters or functions. Traditionally, the bound is applied to unbiased estimators of unknown parameters determining a state, but it applies just as well to functions of the state. Let $R = \braket{M}$ be some function of a state which can be written as the expectation value of some observable $M$, where $M$ has no first-order dependence on $\rho$. The quantum CR bound tells us that for any parameter $x$ of the state,
\begin{align}
\label{eq:BG-CR}
    \Var(M) \geq \frac{\abs{\partial_x R}^2}{I_x},
\end{align}
for variance $\Var(A) = \braket{\delta A^2}$ with $\delta A \coloneqq A - \langle A \rangle$~\cite{braunsteinStatisticalDistanceGeometry1994, liu2019quantum}. When $\logder_x$ is a log-derivative of the state with respect to $x$, $I_x := \Var(\logder_x)$ is the quantum Fisher Information about $x$, which quantifies the rate at which the system's state changes with changes in $x$. Usually one takes a symmetric log-derivative satisfying $\partial_x\rho = \frac{1}{2}\{\logder_x,\rho\}$, but the bound also holds for left-log-derivative satisfying $\partial_x\rho = \logder_x^L\rho$, where in this case $I_x^L = \braket{\logder_x^{L \dag} \logder_x^L}$. The bound is most commonly used for the case $R = \hat x$, an unbiased estimator of parameter $x$, so that $\partial_x R = 1$, but holds in any case where we can calculate (or lower bound) $\partial_x R$. The CR bound is straightforward to derive:
\begin{align}
\begin{split}
    \partial_x \braket{M} &= \tr{M\partial_x\rho} = \frac{1}{2}\tr{M\{\logder_x,\rho\}}\\
    &= \frac{1}{2}\braket{\{M,\logder_x\}} = \frac{1}{2}\braket{\{\delta M,\delta\logder_x\}},
\end{split}
\end{align}
where $\braket{\logder_x} = \tr{\partial_x\rho}=0$. The Cauchy-Schwarz inequality guarantees that
\begin{align}
\begin{split}
    \abs{\partial_x R}^2 &= \abs{\frac{1}{2}\braket{\{\delta M,\delta\logder_x\}}}^2 = \abs{\Cov(M,\logder_x)}^2\\
    &\leq \Var(M)\Var(\logder_x),
\end{split}
\end{align}
for covariance $\Cov(A,B) \coloneqq \langle \{A,B\} \rangle/2 - \langle A \rangle \langle B \rangle = \braket{\frac{1}{2}\{\delta A, \delta B\}}$. This can be rearranged to get the lower bound on $\Var(M)$ in Eq.~\eqref{eq:BG-CR}. 

When multiple $M$ satisfy $\rob = \braket{M}$, the CR bound applies to them all. Moreover, the CR bound is saturable by picking $M$ such that $\delta M \propto \logder_x$. There is guaranteed to be a satisfying $M$ which has no first-order dependence on $x$; however it can potentially have dependence on other parameters or on the state itself. 
%, so here we do not assume the CR bounds are saturable.

\subsection{Danskin's Theorem}
\label{sec:bg-danskin}
Danskin's theorem gives a formula for the derivative of a function which is a maximum over a compact set, such as resource robustness. 
Let $Z \subset\reals^m$ be a compact subset, $g:\ \reals^m\cross Z\rightarrow\reals$ continuous, and $g(\cdot, z):\ \reals^m\rightarrow\reals$ convex for every $z\in Z$. Define the function $f:\ \reals^m\rightarrow\reals$
\begin{align}
    f(x) = \max_{z\in Z} g(x,z),
\end{align}
and $Z_x$ the set of maximizing points
\begin{align}
    Z_x = \{z\ \vert\ g(x,z) = f(x)\}.
\end{align}

Then Danskin's theorem~\cite{bertsekas1999nonlinear} says that
\begin{enumerate}
    \item If $\abs{Z_x}=1$ so that the optimizing $z^*$ is unique, then the derivative of $f$ is well-defined as the derivative of $g$ in its first argument. That is,
    \begin{align}
        \nabla f(x) = \nabla_x g(x,z^*).
    \end{align}
    \item If $\abs{Z_x}>1$, then $f$ is differentiable iff $\nabla_x g(x,z^*)$ is the same for every $z^* \in Z_x$. When $f$ is not differentiable, it instead has a one-sided directional derivative in each direction $y$, given by
    \begin{align}
        \partial_y f(x) = \max_{z\in Z_x} \partial_y g(x,z),
    \end{align}
    where $\partial_y g(x,z)$ is the directional derivative of $g(\cdot,z)$ in the direction $y$.
\end{enumerate}

In the contexts used here, $f$ will be either standard or generalized resource robustness, both of which can be expressed in the form necessary for Danskin's theorem, either by Hermitian matrices in $\mathbb{C}^{2^n\cross 2^n}$, or in terms of a vector of real parameters in $\reals^{4^n}$, for example their real and imaginary coefficients. In the case of robustness $Z\sim\mathcal{W}$, i.e., $Z$ is (isomorphic to) the convex set of witnesses. 

\subsubsection{Danskin's Theorem for Robustness Witnesses}
\label{sec:danskin_robustness}
\begin{comment}
In the context of robustness, for example, we take $\rho_x$ as a state defined in terms of a parameter vector $x$ and $W$ defined in terms of its coefficients, so that $g(x,W) = \tr{\rho_x W}$ and $f(x) = \rob(\rho_x)$, for either standard or generalized robustness [\cref{eq:robustness_dual} or \cref{eq:SDP}]. Then Danskin's theorem guarantees that for each direction $y$ in $\reals^{4^n}$, the one-sided directional derivatives are
\begin{align}
    \partial_y \rob(\rho) = \max_{W\in\mathcal{W}_x}\tr{W\partial_y\rho},
\end{align}
where $\mathcal{W}_x\subseteq\mathcal{W}$ is the set of $W\in\mathcal{W}$ maximizing $\tr{W\rho_x}$, i.e., the set of optimal witnesses. Moreover, if $\abs{\mathcal{W}_x}=1$ so that $\mathcal{W}_x$ has a single element $W^*$, then the gradient in $\reals^{4^n}$ is
\begin{align}
    \nabla \rob(\rho) = \tr{W^*\nabla_x\rho_x},
\end{align}
or equivalently
\begin{align}
    \partial_y \rob(\rho_x) = \tr{W^*\partial_y\rho_x}\ \ \ \ \ \forall y.
\end{align}

In the case where $\rho_x$ depends on a single parameter $x$ and $\rob(\rho_x)$ is differentiable while $\abs{\mathcal{W}_x}>1$, and extension of Danskin's theorem in Appendix~\ref{app:sampling_degenerate} shows that

\begin{align}
    \frac{d}{dx}\rob(\rho_x) = \tr{W_x\partial_x\rho_x}\ \ \ \ \ \forall W_x\in\mathcal{W}_x,
\end{align}

i.e., the derivative can be written equivalently in terms of any optimal witness at $x$. 
\end{comment}

In the context of robustness used in this paper, we take $\rho_x$ defined in terms of a single parameter $x$ and $W$ defined in terms of its coefficients, so that $g(x,W) = \tr{\rho_x W}$ and $f(x) = \rob(\rho_x)$, for either standard or generalized robustness [\cref{eq:robustness_dual} or \cref{eq:SDP}]. Let $\mathcal{W}_x\subseteq\mathcal{W}$ be the set of $W\in\mathcal{W}$ maximizing $\tr{W\rho_x}$, i.e., the set of optimal witnesses at $x$. Then Danskin's theorem guarantees that when $\rob(\rho_x)$ is a differentiable function of $x$, for example when $\abs{\mathcal{W}_x}=1$,
\begin{align}
    \frac{d}{dx}\rob(\rho_x) = \tr{W_x\partial_x\rho_x}\ \ \ \ \ \forall W_x\in\mathcal{W}_x,
\end{align}
i.e., the derivative can be written equivalently in terms of any optimal witness at $x$. While Danskin's theorem gives other results when $\rob(\rho_x)$ is non-differentiable, we will only use the differentiable result here.

Compactness of $\mathcal{W}$, required for Danskin's theorem, is guaranteed whenever robustness has a finite upper bound $\rob_\mathrm{max}$ for the relevant finite-dimensional state space, as in this case we can bound $\Vert W \Vert \leq \rob_\mathrm{max}\ \ \forall W\in\mathcal{W}$. Let $\lambda_\mathrm{max},\lambda_\mathrm{min}$ be the largest and smallest eigenvalues of $W$, with eigenstates $\psi_\mathrm{max},\psi_\mathrm{min}$. Then $\lambda_\mathrm{max} = \tr{W \psi_\mathrm{max}}\leq\rob(\psi_\mathrm{max})\leq \rob_\mathrm{max}$. For generalized robustness, $W\geq 0$, so $\Vert W \Vert = \lambda_\mathrm{max}$. For standard robustness, $-W$ is also a valid witness, so $-\lambda_\mathrm{min} \leq \tr{-W\psi_\mathrm{min}} \leq \rob_\mathrm{max}$. In either case, $\Vert W \Vert = \max\{\lambda_\mathrm{max},-\lambda_\mathrm{min}\}\leq\rob_\mathrm{max}$.  $\mathcal{W}$ is also guaranteed compact when free states span the full operator space, or, for generalized robustness, whenever $\free$ contains any full rank state (such as the maximally mixed state). 

\section{Resource Cram\'er-Rao Bounds}
\label{sec:CR-bounds}

Say our goal is to estimate some quantity, in this case resource robustness $\rob$, which can be written as the expectation value of an observable $\rob = \braket{W}$. For an unbiased estimate $\hat\rob$ resulting from averaging $m$ independent measurements of $W$, the mean squared error $\epsilon^2$ of the estimate is related to the variance of $\hat\rob$~\cite{tsang2020quantum} by 
\begin{align}
    \epsilon^2 = \mathbb{E}[(\hat\rob-\rob)^2] = \Var(\hat\rob) = \frac{\Var(W)}{m}.
\end{align}

To estimate $\rob$ to target error $\epsilon$ then requires $m = \Var(W)/\epsilon^2$ samples. When $\Var(W)\sim \exp(n)$, we require $m\sim \exp(n)$ samples in order to accurately estimate the true value $\rob$, making its measurement inefficient. If our goal is instead to estimate $\rob$ to low multiplicative error $\epsilon' = \epsilon/\rob$, the required number of samples goes as $m = (\Var(W)/\rob^2)/\epsilon'^2$. That is, it now depends on the relative variance $\Var(W)/\rob^2$. By using Cram\'er-Rao (CR) bounds to lower bound $\Var(W)$, we can then determine cases where $\Var(W)$ or $\Var(W)/\rob^2$ are large, making $\rob$ inefficient to estimate to low absolute or multiplicative error from standard measurements, respectively.

Resource Robustness measures are a natural context for CR bounds, as their dual definition allows us to write them as the expectation value $\rob = \braket{W}$ for a (generally non-unique) optimal resource witness $W$. 
%Consider applying a perturbation $\delta\rho$ of strength $c$ to a target state $\rho$, creating the perturbed state $\rho_c := \rho + c\delta\rho$. 
We will mainly consider perturbative CR bounds constructed as follows. We will pick some initial state $\rho$ with no parameter dependence, and perturb it to $\rho_c := \rho + c\delta\rho$, where $\tr{\delta\rho}=0$, then calculate the CR bound in the limit $c\rightarrow0$. 
%Let $\rob_W(\rho_c) := \tr{W\rho_c}$ for any optimal witness $W$ for the state $\rho$. 
While $\partial_c\rob(\rho_c) \neq \partial_c \rob_W(\rho_c)$ in general, we have $\rob_W(\rho_c)\big\vert_{c=0} = \rob(\rho)$, and calculating $\partial_c\rob_W(\rho_c)\big\vert_{c=0}$ will turn out to be sufficient to place a CR bound on the variance of $W$ in the state $\rho$.

As an alternative approach, Danskin's theorem [\cref{sec:bg-danskin}] guarantees that $\partial_c\rob(\rho_c)\big\vert_{c=0} = \tr{W \partial_c \rho_c} \big\vert_{c=0}$ for any optimal $W$, as long as $\rob(\rho_c)$ is differentiable at $c=0$. The differentiability condition occurs, for example, when $W$ is the unique optimal witness for $\rho$. The bounds in this section are framed in terms of $\partial_c\rob_W$ for generality, but hold equally well for $\partial_c\rob$ when $\rob$ is differentiable. In the latter case, they lower bound the variance of any unbiased estimator of $\rob$, not just that of an optimal witness.

%While we could apply Danskin's theorem to find the directional derivative $\partial_c\rob(\rho_c)$, this is unnecessarily restrictive, as in the case of multiple optimal witnesses the resulting CR bound would apply only to the witness which maximizes the directional derivative. Instead, define $\rob_W(\rho_c) := \tr{W\rho_c}$ for any optimal witness $W$ for the state $\rho$, which is now held fixed. While $\partial_c\rob(\rho_c) \neq \partial_c \rob_W(\rho_c)$ in general, we have $\rob_W(\rho_c)\big\vert_{c=0} = \rob(\rho)$, and calculating $\partial_c\rob_W(\rho_c)\big\vert_{c=0}$ will be sufficient to bound the variance of $W$ in the state $\rho$.

In this section, we will use this technique to bound the variance of arbitrary optimal witnesses in two cases: robustness and generalized robustness for mixed states, and generalized robustness of pure states. In Appendix~\ref{app:CR-moments}, we also bound the variance of observables which can be used to estimate moments, including stabilizer, purity, and entanglement moments. The bounds presented in this section are not necessarily saturable, but are instead intended to be general and interpretable. Tighter CR bounds can be derived in terms of different state perturbations, but incur dependence on the specific $W$ or on diagonalization of $\rho$, which limits their utility. 

\subsection{Robustness of Mixed States}

The first CR bound we find is on the variance of an optimal witness of robustness or generalized robustness of sufficiently mixed states. 
Specifically, this holds for mixed states for which $\mathcal{B}(\mathcal{V}_\rho)\cap\free\neq\varnothing$, where $\mathcal{V}_\rho$ is the subspace spanned by the eigenvectors of $\rho$ with nonzero eigenvalues. This class of states includes full-rank states as a subset.

To place bounds on the variance of a robustness estimator at $\rho$, we consider a derivative of the perturbed state $\rho_c = \rho + c(\rho - \phi)$ in the limit $c\rightarrow0$, where $\phi \in \mathcal{B}(\mathcal{V}_\rho)\cap\free\neq\varnothing$ can be any free state composed of eigenvectors of $\rho$, and consider the fixed witness ``robustness'' $\rob_W(\rho_c)$. 

Let $\rob$ denote either standard or generalized robustness, and let $W$ be an optimal witness so that $\rob(\rho) = \tr{W\rho} = \rob_W(\rho)$ and $\abs{\tr{W\phi'}}\leq1$ for all free states $\phi'$.
Then, it holds that
\begin{align}
\abs{\partial_c\rob_W(\rho_c)}\big\vert_{c=0} &= \abs{\tr{W\partial_c\rho_c}}\big\vert_{c=0} = \abs{\rob(\rho)-\tr{W\phi}} \nonumber \\
    &\geq \abs{\rob(\rho)-1},
\end{align}
since $\rob(\rho) - \tr{W\phi} \geq \rob(\rho)-1 \geq 0$.

In theory, one could find a symmetric log-derivative generating $\rho_c$, but its form would depend explicitly on the diagonalization of $\rho$ and not be particularly enlightening. Instead, we consider $\rho_c$ as being generated by the left log-derivative $\logder_c^L = (\rho-\phi)\rho^{-1}$, where $\rho^{-1}$ is the inverse of $\rho$ within $\mathcal{V}_\rho$, which is always well-defined. Then the left Fisher information in the $c\rightarrow0$ limit is
\begin{align}
    I_c^L\big\vert_{c=0} &= \Var(\logder_c^L)\big\vert_{c=0} = \braket{\logder_c^{L\dag}\logder_c^L}\big\vert_{c=0} = \tr{\rho^{-1}(\rho-\phi)^2}\nonumber\\
    &= \tr{\rho^{-1}\phi^2}-1.
\end{align}
Combining these terms with the CR formula in \cref{eq:BG-CR}, we obtain:

\begin{theorem}
\label{thm:CR_mixed}
For any state $\rho$ and any optimal witness $W$ of its standard or generalized robustness $\rob$, the variance of $W$ obeys
\begin{align}
    \Var(W) \geq \frac{(\rob-1)^2}{\tr{\rho^{-1}\phi^2}-1}
\end{align}
for every free state $\phi$ within the span of the eigenvectors of $\rho$ with nonzero eigenvalues. 
\end{theorem}

The denominator can also be written as $\Vert \rho^{-\frac{1}{2}}(\rho-\phi)\Vert_2^2$, highlighting that it is always non-negative, and only zero when $\rho=\phi$. When multiple valid $\phi$ exist, the tightest bound comes from picking the $\phi$ which minimizes $\tr{\rho^{-1}\phi^{2}}$, which will tend to be one with more support on the eigenspaces of $\rho$ corresponding to larger eigenvalues.

As the bound's explicitly dependence on $\rob$ is quadratic, the lower bound on the relative variance, $\Var(W)/\rob^2$, has no direct dependence on $\rob$ and only on the overlap term $\tr{\rho^{-1}\phi^{2}}$. Additionally, since any resourceful state has $\rob > 1$, this bound shows that the variance cannot be zero except in the limit where the denominator approaches infinity, which is not possible when $\phi\in\mathcal{B}(\mathcal{V}_\rho)$. 
%In order to reduce the relative variance, defined as $\Var(W)/\rob^2$, below a threshold one would need to find a state minimizing 
Thus, a necessary condition for measuring mixed-state robustness with zero variance is that $\mathcal{B}(\mathcal{V}_\rho)\cap\free=\varnothing$, i.e. there is no free state within the span of the nonzero eigenvectors of $\rho$, so that the bound breaks down. This occurs for any resourceful pure state, so we consider these separately in the next subsection.

This bound is greater for highly mixed states. Consider a nearly-pure state $\rho = (1-\epsilon)\ketbra{\psi_0}{\psi_0}+\epsilon\ketbra{\psi_1}{\psi_1}$, and some $\phi = \ketbra{\phi}{\phi}$ in the relevant subspace with $\braket{\phi\vert\psi_1}\neq0$. In this case the denominator becomes $(1-\epsilon)^{-1}\abs{\braket{\psi_0\vert\phi}}^2 + \epsilon^{-1}\abs{\braket{\psi_1\vert\phi}}^2 -1 \in \Omega(\epsilon^{-1})$, which becomes infinite as $\epsilon\to0$, causing the lower bound to approach zero when $\rob$ is held fixed. In contrast, the bound is tight for a number of classes of mixed states, with the example of robustness of entanglement for isotropic states considered in Appendix \ref{app:scope_mixed}.

%This bound is tighter for highly mixed states. For a nearly-pure state $\rho$ with minimum nonzero eigenvalue $p_0$, the denominator can be bounded by $\tr{\rho^{-1}\phi^2}-1 \leq p_0^{-1} - 1$, giving $\Var(W) \geq (\rob-1)^2/(p_0^{-1}-1)\sim p_0 (\rob-1)^2$ for small $p_0$. Thus the bound approaches zero as $\rho$ approaches purity, even when $\rob$ is held fixed. In contrast, the bound is tight for a number of classes of mixed states, with the example of robustness of entanglement for isotropic states considered in Appendix \ref{app:scope_mixed}.

\subsection{Generalized Robustness of Pure States}

Let us consider the generalized robustness $\rob^+$ of a pure state $\ket{\psi}$. In this case, we consider the perturbation $\ket{\psi_c} = \ket{\psi} + c\ket{\psi'}$, where 
%$\ket{\psi'} = \braket{\psi\vert\phi}\ket{\psi} - \ket{\phi}$ is the projection of $-\ket{\phi}$ onto the orthogonal subspace to $\ket{\psi}$. Pick $\ket{\phi}$ as the free state with maximal overlap with $\ket{\psi}$, that is, $\abs{\braket{\phi\vert\psi}} = \max_{\ket{\phi'}}\abs{\braket{\phi'\vert\psi}} = \sqrt{F}$, where $F$ is the free-state fidelity of $\ket{\psi}$. 
$\ket{\psi'} = (F - \phi)\ket{\psi}$, $F$ is the free-state fidelity of $\ket{\psi}$, and $\phi$ is a (potentially mixed) free state with maximal fidelity to $\ket{\psi}$, so that $F = \braket{\psi\vert\phi\vert\psi}$. 
The CR bound derivation is slightly more involved than the mixed state case, but no restriction must be placed on $\ket{\psi}$ -- the bound holds for every resourceful pure state. Under this construction, in Appendix~\ref{app:CR-proof-GR} we show the following result:

\begin{theorem}
\label{thm:CR_pure}
For any pure state with free-state fidelity $F$ and generalized robustness $\rob^+$, the variance of any optimal witness $W$ of $\rob^+$ obeys
\begin{align}
    \Var(W) &\geq \frac{\rob^+\left(\sqrt{\rob^+F} - 1\right)^2}{1-F}.
\end{align}
\end{theorem}

Unlike the mixed-state case, this bound can be zero for even resourceful states, in the case where $\rob^+ = 1/F$. This case describes states which are proportional to their own optimal witness, and can thus be measured with zero variance by measuring the state itself. A common example is the $\ket{T}$ state for generalized robustness of magic. Upper bounding $1-F \leq 1$ and inspecting the relative variance reveals
\begin{align}
\label{eq:CR_pure_relative}
    \frac{\Var(W)}{\rob^{+2}} \geq \left(\sqrt{F}-\frac{1}{\sqrt{\rob^+}}\right)^2, 
\end{align}
which grows larger the further $\rob^+$ is from its lower bound $1/F$. Increasing both $\rob^+$ and $1/F$ by some factor $c$ reduces the relative variance bound by a factor of $1/c$, suggesting that generalized robustness will tend to be harder to measure with low relative error for less resourceful pure states. In particular, if $\rob^+$ and $1/F$ are both exponential in system size while maintaining their ratio, the relative error bound is exponentially small in system size. \cref{thm:CR_mixed} does not have the same behavior, suggesting that generalized robustness measures on highly resourceful pure states may be better candidates for estimating the capabilities of quantum devices.

A comparison of the bound in \cref{thm:CR_pure} and those in the next section for low- and high-resource states is considered in Appendix \ref{app:scope_pure}, including a family of states for which all bounds are tight. 

In Appendix~\ref{app:CR-moments}, we extend the present approach based on CR bounds to the context of moments, which includes stabilizer, purity, and entanglement moments as example cases. While nonlinear in the state, these quantities can be estimated from an observable on $k$ copies of a state, of which we bound the variance. 

\subsection{Upper Bounds}

When either notion of robustness takes a maximum value $\rob_\mathrm{max}$ over the state space, we can also place a loose upper bound on the variance of any witness. As mentioned in \cref{sec:danskin_robustness}, in this case $\Vert W \Vert \leq \rob_\mathrm{max}$ for any witness $W$. For standard robustness, this implies $-\rob_\mathrm{max} \mathbb{I} \leq W \leq \rob_\mathrm{max} \mathbb{I}$, so 
\begin{align}
    \Var(W) = \braket{W^2}-\braket{W}^2\leq \rob_\mathrm{max}^2 - \rob^2,
\end{align}
while for generalized robustness, $0 \leq W \leq \rob^+_\mathrm{max}\mathbb{I}$, and 
\begin{align}
    \Var(W) \leq \rob^+(\rob^+_\mathrm{max} - \rob^+).
\end{align}
So as $\rob \to \rob_\mathrm{max}$ in either case, the witness variance necessarily approaches zero. Moreover, in both cases the relative variance upper bounds depend only on $\rob_\mathrm{max}/\rob$, and has no inherent scaling with $\rob$, unlike the absolute variance. 

\subsection{Estimating Robustness with Bounded Variance}
\label{sec:achievable_witnesses}

The derivations of Theorems \ref{thm:CR_mixed} and \ref{thm:CR_pure} do not require witness optimality; they apply equally well for any witness $W$, with the caveat that \cref{thm:CR_pure} requires $\braket{W} \geq 1/F(\psi)$. As we will see shortly, this imposes no restriction here. Measurement of sub-optimal witnesses yields a lower bound on robustness, so this motivates asking how good an optimal witness can be achieved within a given variance threshold. Equivalently, how good an estimate of robustness can be achieved with a fixed number of measurements. Substituting $\rob\to\braket{W}$ into \cref{thm:CR_mixed} and solving for it gives the immediate bound
\begin{align}
    \braket{W} \leq \min \left\{1+\alpha \sqrt{\Var(W)}, \rob(\rho) \right\},
\end{align}
where $\alpha = \sqrt{\tr{\rho^{-1}\phi^2}-1}$ depends on $\rho$ but not on $W$, and $\rob$ stands for either standard or generalized robustness. When relative variance $V_\mathrm{rel}$ is held fixed instead, the resulting bound instead goes as
\begin{align}
    \braket{W} \leq \min \left\{\frac{1}{1-\alpha\sqrt{V_\mathrm{rel}}},\rob(\rho)\right\},
\end{align}
 for $\alpha\sqrt{V_\mathrm{rel}}<1$, and the trivial $\braket{W} \leq \rob(\rho)$ otherwise. This behaves similarly to the absolute variance bound for small $V_\mathrm{rel}$ but yields much higher allowable $\braket{W}$ for large $V_\mathrm{rel}$, achieving the true value $\braket{W}=\rob$ for some $V_\mathrm{rel}^* \leq \alpha^{-1/2}$. 

For generalized robustness of a pure state, the sub-optimal witness $W_\psi = \psi/F(\psi)$ can always be measured with zero variance, and gives $\braket{W_\psi} = 1/F(\psi)$, giving a somewhat trivial lower bound. To improve on this, we then only need to consider witnesses with $\braket{W} > 1/F(\psi)$, which satisfies the requirement in \cref{thm:CR_pure}. This allows us to substitute $\rob^+ \to \braket{W}$ in the associated bound. The bound can be solved for $\rob^+$, but the resulting form is opaque. It is more transparent to instead solve for the best achievable $\braket{W}$ with fixed relative variance, which gives
\begin{align}
    \braket{W} \leq \min \left\{\frac{1}{F}\frac{1}{(1-\beta\sqrt{V_\mathrm{rel}})^2}, \rob^+(\psi) \right\}
\end{align}
when $\beta\sqrt{V_\mathrm{rel}} < 1$, where $\beta = \sqrt{\frac{1-F}{F}}$ is independent of $W$. At fixed $F$ this offers a quadratic improvement over the mixed-state bound. For small relative variance, we see that the max achievable robustness bound goes as $\braket{W} \lesssim 1+\alpha\sqrt{V_\mathrm{rel}}$ while for generalized robustness it goes as $\braket{W} \lesssim \frac{1}{F} + \frac{2\beta}{F}\sqrt{V_\mathrm{rel}}$.

\section{Generalized Robustness Witness Geometry}
\label{sec:geometry}

In the previous section we saw that the pure-state variance of an optimal generalized robustness witness depends on the relationship between the state's generalized robustness $\rob^+$ and free-state fidelity $F$. In fact, for pure states, the variance can be written exactly in terms of the angle between the target state and a witness state, making the variance an inherently geometric quantity. In this section, we will use this geometric nature to get a different lower bound on witness variance. Its geometric origin makes it easier to interpret and potentially to generalize, but at present only holds for states with high free-state fidelity, typically those with low resource. 

For a pure state $\ket{\psi}$, there always exists an optimal generalized robustness witness which is proportional to a pure state projector~\cite{Regula2018Convex, takagiPRL, johnston2018evaluating}:
\begin{align}
\label{eq:OptimalWitnessPure}
W = \frac{\ketbra{\omega}{\omega}}{F(\omega)}, 
\end{align}
where $\ket{\omega}$ is a normalized state and $F(\omega)$ is the free-state fidelity of state $\ket{\omega}$. 

One can also construct such a pure optimal witness $W$ from any optimal witness $W'$ by setting
\begin{align}
\label{eq:witness-purification}
    W = \frac{W'\ketbra{\psi}{\psi}W'}{\rob^+(\psi)}.
\end{align}
The resulting $W$ obeys $\braket{\phi\vert W\vert \phi} = \abs{\braket{\phi\vert W'\vert\psi}}^2/\rob^+ \leq \braket{\phi\vert W'\vert\phi}\leq 1\ \forall\phi\in\free$ by Cauchy-Schwarz, verifying that it is a witness. Moreover, since $W\ket{\psi} = W'\ket{\psi}$, it is both optimal and has $\Var(W) = \Var(W')$. Thus, while we work with pure optimal witnesses in this section, the variance results hold for any optimal witness. However, unlike the previous section, they do not extend to other unbiased estimators of $\rob^+$. 

For pure $W$, the variance of the witness operator takes a geometric form:
\begin{align}
\label{eq:var_bound_derivation_geometric}
    \Var(W) &=  \braket{W^2} - \braket{W}^2= \frac{\abs{\braket{\psi\vert\omega}}^2}{F(\omega)^2} - \frac{\abs{\braket{\psi\vert\omega}}^4}{F(\omega)^2}\nonumber\\
    &= (\rob^+)^2\left(\frac{1-\abs{\braket{\psi\vert\omega}}^2}{\abs{\braket{\psi\vert\omega}}^2}\right).
\end{align}
Make the substitution $\theta_{\psi\omega} = \cos^{-1}(\abs{\braket{\psi\vert\omega}})$ representing the state-space angle between states $\ket{\psi}$ and $\ket{\omega}$. Since $\sin^2(\theta_{\psi\omega}) = 1-\abs{\braket{\psi\vert\omega}}^2$, the variance trigonometrically simplifies to
\begin{align}
\label{eq:varianceW}
    \Var(W) = (\rob^+)^2 \tan^2(\theta_{\psi\omega}).
\end{align}
For pure states, the variance of the optimal witness is quadratic in generalized robustness but also depends on the alignment between the original and witness states. If one can pin down the set of potential witness states, one can lower bound the variance in terms of the distance between the original state and this set. The tighter the set of potential witnesses, the greater the minimum possible angle $\theta_{\psi\omega}$, and the tighter the variance lower bound.

\subsection{Witness Free-State Fidelity}

As a first step towards characterizing this set, consider the witness free-state fidelity $F(\omega)$. We can upper bound $F(\omega)$ in terms of $F(\psi)$ as 
\begin{align}
    \begin{split}
        F(\psi) &\geq \frac{1}{\rob^+(\psi)} = \frac{F(\omega)}{\abs{\braket{\psi\vert\omega}}^2} \geq F(\omega),
    \end{split}
\end{align}
with the two having equality iff $\ket{\psi}=\ket{\omega}$, which also implies $\rob^+(\psi) = 1/F(\psi)$. 

When $\free$ is the convex hull of pure free states $\free_\mathrm{pure}$ and $\ket{\psi}$ is a pure state, 
%For a pure state $\ket{\psi}$, 
$\rob^+(\psi)$ admits a primal definition as the pure-state extent~\cite{seddon2021quantifying, Regula2018Convex, bravyi2019simulation}
\begin{align}
\label{eq:purestate-extent}
    \rob^+(\psi) = \xi(\psi) = \min_{\vec c} \left\{\Vert \vec c\Vert_1^2\ \Big\vert\ \ket{\psi} = \sum_{i=1}^r c_i\ket{\phi_i} \right\},
\end{align}
where $c_i \in\mathbb{C}$ are nonzero, $\ket{\phi_i}\in\free_\mathrm{pure}$, and $\Vert \vec c\Vert_1^2 = (\sum_i \abs{c_i})^2$. 

For a given decomposition, call a free state ``active'' if it has a nonzero coefficient. When $\ket{\psi}$ admits an optimal decomposition with minimum angle $\cos(\theta_f) = \max_{i\neq j}\{\abs{\braket{\phi_i\vert\phi_j}}\}$ between active free states, we can also place a $\ket{\psi}$-independent upper bound on $F(\omega)$. This is guaranteed to exist in resource theories with a minimal angle between pure free states, such as magic ($\theta_f=\frac{\pi}{4}$~\cite{bravyi2019simulation}) and coherence ($\theta_f = \frac{\pi}{2}$). It also holds for entanglement with $\theta_f = \frac{\pi}{2}$, as in this case the Schmidt decomposition is optimal, even though general free states have no minimal separating angle. 

\begin{lemma}
\label{lem:witness_fidelity}
When a state admits an optimal free-state extent decomposition with $r$ active free states and minimum angle $\theta_f$ between active free states, any pure optimal witness state $\ket{\omega}$ must have 
\begin{align}
\label{eq:witness_fidelity_bound}
F(\omega) \leq F_r:=\frac{1+(r-1)\cos(\theta_f)}{r}.
\end{align}
For any resourceful state, $r \geq 2$, so $F(\omega) \leq F_2 = (1+\cos(\theta_f))/2 = \cos^2(\theta_f/2)$. 
\end{lemma}
\begin{proof}
Plugging an optimal free-state decomposition of $\ket{\psi}$ into the dual definition $\rob^+(\psi) = \abs{\braket{\psi\vert\omega}}^2/F(\omega)$ and equating $\xi(\psi) = \rob^+(\psi)$ reveals that $\abs{\braket{ \omega\vert \phi_i}}^2 = F(\omega)$ for every free state $\ket{\phi_i}$ in the sum~\cite{seddon2021quantifying}.

Pick the phase of each $\ket{\phi_i}$ so that $\braket{\omega\vert\phi_i} = \sqrt{F(\omega)}$, and set $\ket{v} = \sum_i \ket{\phi_i}$. Then $\braket{\omega\vert v} = r\sqrt{F(\omega)}$. Since $\braket{\phi_i\vert\phi_j}\leq\cos(\theta_f)$, $\braket{v\vert\phi_i} \leq 1+(r-1)\cos(\theta_f)$, and we can bound
\begin{align}
\begin{split}
    r^2F(\omega) &= \braket{\omega\vert v}^2 \leq \braket{v\vert v} = \sum_{i} \braket{v\vert\phi_i}\\ 
    &\leq r(1+(r-1)\cos(\theta_f)).
    \end{split}
\end{align}
Solving for $F(\omega)$ then shows \cref{eq:witness_fidelity_bound}.
\end{proof}

Note that $r$ can be lower-bounded by the state's free-state rank. Plugging $F_r \geq F(\omega)$ into the second line of \cref{eq:var_bound_derivation_geometric} gives:

\begin{theorem}
\label{thm:var_bound_r-R}
    For a pure state $\ket{\psi}$, the variance of any optimal generalized robustness witness $W$ satisfies
    \begin{align}
        \Var(W) \geq \rob^+\left(\frac{1}{F_r}-\rob^+\right),
    \end{align}
    where $F_r := (1+(r-1)\cos(\theta_f))/r\leq \cos^2(\theta_f/2)$, $r$ is the number of active free states in the optimal free-state extent decomposition of $\ket{\psi}$, and $\theta_f$ is the minimal angle between active free states. 
\end{theorem}

\begin{comment}
\begin{align}
    \Var(W) \geq \rob^+\left(\frac{1}{F_r}-\rob^+\right).
\end{align}
\end{comment}

In the $r=2$ case, the geometric interpretation of the lemma is particularly simple: $\ket{\omega}$ is equidistant to $\ket{\phi_1},\ket{\phi_2}$, so $\theta_{ \omega \phi_1} = \theta_{ \omega \phi_2}$. As the angle between the two free states is at least $\theta_f$, the triangle inequality gives
\begin{align}
    \begin{split}
        \theta_f &\leq \theta_{\phi_1 \phi_2} \leq \theta_{ \omega \phi_1} + \theta_{ \omega \phi_2}= 2\theta_{ \omega \phi_1}.
    \end{split}
\end{align}
Thus, $F( \omega) = \cos^2(\theta_{ \omega \phi_1}) \leq \cos^2(\theta_f/2)$, matching the general bound in the $r=2$ case. For $r > 2$, $\ket{\omega}$ lies at the center of an $r$-vertex polytope. 

When $\theta_f = \frac{\pi}{2}$ and the witness $W^* = \ketbra{v}{v} = \sum_{ij}\ketbra{\phi_i}{\phi_j}$ is feasible, $W^*$ is both optimal and saturates the variance bound in \cref{thm:var_bound_r-R}. This is guaranteed when all pure free states are orthogonal, such as for coherence, where they are basis states~\cite{johnston2018evaluating}. It also occurs for bipartite entanglement, where the active free states are the product states resulting from the Schmidt decomposition, which can be shown to obtain the optimal pure-state extent~\cite{johnston2018evaluating}. Feasibility of $W^*$ then follows from a Cauchy-Schwarz argument showing $\abs{\braket{v\vert a,b}}\leq 1$ for any product state $\ket{a,b}$; see, e.g., \cref{eq:separable-cauchy}.

%\cref{thm:var_bound_r-R} is saturable in the case where $\theta_f = \frac{\pi}{2}$, where it is achieved by the optimal witness $W^* = \ketbra{v}{v} = \sum_{ij}\ketbra{\phi_i}{\phi_j}$. This is the case for the generalized robustnesses of coherence (where $\ket{\phi_i}$ are basis states) and of bipartite entanglement (where $\ket{\phi_i}$ are Schmidt product states), and results in the bound equaling the true variance $\Var(W^*) = \rob^+(r-\rob^+)$~\cite{johnston2018evaluating}. While the free states of entanglement are in general arbitrary product states, which can be close together, $\theta_f = \frac{\pi}{2}$ is achieved by the Schmidt decomposition, which achieves the optimal pure-state extent value~\cite{johnston2018evaluating}. This also implies that $W^*$, written in the Schmidt basis, is an optimal witness for bipartite entanglement.

In contrast, when $\cos(\theta_f) > 0$, one can lower bound $F_r \geq \cos(\theta_f)$, so \cref{thm:var_bound_r-R} is only nontrivial when $\rob^+ < \sec(\theta_f)$. For example, in the case of magic, the bound is only of interest for $\rob^+ < \sqrt{2}$. Within this regime it is still useful for bounding the relative variance of optimal witnesses. For example, for a nearly-free state with $\rob^+ = 1+\epsilon$ and $r=2$, $F_2 = \cos^2(\frac{\theta_f}{2})$, and the bound yields
\begin{align}
        \frac{\Var(W)}{(\rob^+)^2} &\geq \frac{1}{F_r\rob^+}-1 = \sec^2(\frac{\theta_f}{2})(1-\epsilon)-1+O(\epsilon^2) \nonumber \\
        &= \tan^2(\frac{\theta_f}{2}) - \sec^2(\frac{\theta_f}{2})\epsilon + O(\epsilon^2).
\end{align}
This highlights that for a sufficiently low-resource non-free state, the optimal witness relative variance can only be zero if $\theta_f \sim O(\sqrt{\epsilon})$, necessitating that $\free$ be continuous as $\epsilon\to0$.

\subsection{Geometric Variance Bound}

We can use \cref{lem:witness_fidelity} and \cref{eq:varianceW} to lower bound the variance of the optimal witness when $F(\psi) \geq F_r$.

\begin{theorem}
\label{thm:var_bound_inst}
    For a pure state $\ket{\psi}$ with $F := F(\psi) > F_r := (1+(r-1)\cos(\theta_f))/r$, the variance of any optimal witness $W$ is lower bounded by
    \begin{align}
    \label{eq:var_bound_inst}
        \mathrm{Var}(W) \geq (\rob^+)^2\left(\frac{F - F_r}{\sqrt{F(1-F)}+\sqrt{F_r(1-F_r)}}\right)^2,
    \end{align}
    where $r$ is the number of active free states in the optimal pure state extent decomposition of $\ket{\psi}$, and $\theta_f$ is the minimal angle between active free states. 
\end{theorem}

\begin{proof}
    We will lower bound $\theta_{\psi \omega}$, then use $\Var(W) = (\rob^+)^2\tan^2(\theta_{\psi\omega})$ and several trig identities. Let $\ket{\phi}$ be any free state maximizing overlap with $\ket{\psi}$, so that $F(\psi) = \cos^2(\theta_{\phi\psi})$. \cref{lem:witness_fidelity} guarantees that $\theta_r := \arccos\sqrt{F_r} \leq \min_{\phi'}\theta_{\phi'\omega}\leq\theta_{\phi\omega}$. Combining this with the triangle inequality gives
    \begin{align}
        \theta_r\leq \theta_{\phi \omega}\leq \theta_{\phi \psi}+\theta_{\psi  \omega}.
    \end{align}
    We can then rearrange terms to solve for $\theta_{\psi\omega}$ to get $\theta_{\psi\omega} \geq \theta_r-\theta_{\phi\psi}$. Plugging this in as a lower bound on $\theta_{\psi\omega}$ in \cref{eq:varianceW} then gives
    \begin{align}
\label{eq:var_bound_tangent}
        \begin{split}
            &\Var(W) \geq (\rob^+)^2\tan^2\left(\theta_r - \theta_{\phi\psi}\right)\\
            &= (\rob^+)^2\tan^2\left(\arccos\sqrt{F_r} - \arccos\sqrt{F}\right).
        \end{split}
    \end{align}
One can then use the trigonometric identities for difference of angles to expand $\sin(\theta_r - \theta_{\phi\psi})$ and $\cos(\theta_r - \theta_{\phi\psi})$, then substitute $\sin(\theta) \to \sqrt{1-\cos^2(\theta)}$, leading to a cancellation of all $\arccos$ terms. The resulting form is lengthy, but multiplying the resulting numerator and denominator by $\sqrt{F(1-F_r)}+\sqrt{F_r(1-F)}$ simplifies it to 
    \begin{align}
        \tan(\theta_r - \theta_{\phi\psi}) = \frac{F - F_r}{\sqrt{F(1-F)}+\sqrt{F_r(1-F_r)}}.
    \end{align}
    Plugging this into \cref{eq:var_bound_tangent} then proves the theorem.
\end{proof}

The denominator in \cref{thm:var_bound_inst} is maximized for $F = F_r = \frac{1}{2}$, so it can also be upper bounded by $1$ to get a simpler but looser bound:
\begin{align}
    \Var(W) \geq (\rob^+)^2(F - F_r)^2.
\end{align}

\cref{thm:var_bound_inst} highlights the approximately-quadratic dependence of relative variance on $F - F_r$, the difference between the state's fidelity and the maximal fidelity allowed for any optimal witness state. $F-F_r$ can be interpreted as the fidelity-distance of $\ket{\psi}$ from the set of possible optimal witness states.
When $\ket{\psi}$ is close to being a free state, the fidelity part of the bound in \cref{thm:var_bound_inst} has a positive relationship with $F(\psi)$. That is, the relative variance lower bound increases as the state becomes less resourceful, highlighting the difficulty of detecting resources in low-resource states. This mirrors the result from the previous section, where relative variance was found to decrease with $\rob^+$, $1/F$, as long as the ratio of the two is held constant.

In Appendix~\ref{app:scope_pure} we provide a comparison of the bounds in this section and the bound in \cref{thm:CR_pure} for low and high-resource states, and discuss a family of states for which all bounds are tight. Unlike in the previous section, the bounds in this section rely on witness optimality, so they cannot be immediately used to  determine the best allowed $\braket{W}$ at fixed variance.

%A comparison of the bounds in this section and the bound in \cref{thm:CR_pure} low and high-resource states is considered in Appendix \ref{app:scope_pure}, including a family of states for which all bounds are tight. Unlike in the previous section, the bounds in this section rely on witness optimality, so they cannot be immediately used to  determine the best allowed $\braket{W}$ at fixed variance.

\section{Sampling hardness of tracking the onset of resource robustness}
\label{sec:Sampling}

An additional challenge emerges if we want to track the onset of resource robustness. That is, if we want to evolve some state and determine when it transitions from being low-resource to exponentially high resource. In this case, even if the start and endpoints of evolution can both be measured with low variance, we show that there always exist parts of the trajectory for which resource robustness cannot be efficiently estimated. While the previous two sections established conditional lower bounds on the variance of optimal witnesses, it is hard to determine when these bounds become exponentially large without additional information ($\phi$ for mixed states, or $F$ for pure states). Such resource-specific information is not needed here; we instead establish global bounds along paths from low to high resource states, for which the integrated variance is guaranteed to be large as long as the path has a local generator. 

Consider a system whose state $\rho_x$ depends on a parameter $x$, ranging from zero to one (without loss of generality), for which $\rob(\rho_x)$ is an absolutely continuous function of $x$, so that $\rob(\rho_x)$ is differentiable almost everywhere. At each point $x$, let $W_x$ be any estimator of $\rob(\rho_x)$, which is locally unbiased in $x$ so that $\partial_x\rob(\rho_x) = \tr{W_x\partial_x\rho_x}$. This includes any optimal witness for $\rho_x$, as shown by Danskin's theorem [\cref{sec:bg-danskin}]. The total change $\Delta\rob:= \rob(\rho_1)-\rob(\rho_0)$ satisfies
\begin{align}
\label{eq:boundchange}
    \left|\Delta\rob \right| &\leq \int_{0}^{1} \left| \partial_x \rob(\rho_x) \right| dx  \nonumber \\ 
    &= \int_{0}^{1}  \left| \tr{W_x\partial_x\rho_x}\right| dx  \nonumber \\ 
    &= \int_{0}^{1}  \left| \Cov(\logder_x, W_x)   \right| dx \nonumber \\ 
    &\leq \int_{0}^{1}  \sqrt{ \Var(\logder_x) \Var(W_x)} dx 
    \nonumber \\ 
    &\leq \sqrt{ \int_{0}^{1}\!\!   \Var(\logder_x) dx } \, \sqrt{ \int_{0}^{1}   \!\! \Var(W_x) dx }.
\end{align}
The third line comes from expanding $\partial_x\rho_x$ in terms of the log-derivative, while the last two lines follow from the Cauchy-Schwarz inequality for operators and functions, respectively. We exclude the measure-zero set of points where $\rob(\rho_x)$ is not differentiable from the integral, as this does not change its total value.  

Equation~\eqref{eq:boundchange} then relates the integrated noise on the estimator to the change in the robustness:

\begin{theorem}
\label{thm:path_bound}
For a path $\rho_0\to\rho_1$ over which $\rob(\rho_x)$ is an absolutely continuous function of $x$, and where $W_x$ is any unbiased estimator of robustness for $\rho_x$ such as an optimal witness, the integrated variance obeys 
\begin{align}
\label{eq:BoundVariance}
    \int_{0}^{1} \Var(W_x) dx \geq \frac{(\Delta\rob)^2}{ \int_{0}^{1} I_x dx },
\end{align}
for quantum Fisher information $I_x = \Var(\logder_x)$ and total change in robustness $\Delta\rob = \rob(\rho_1)-\rob(\rho_0)$.
\end{theorem}
\begin{comment}
\begin{theorem}
\label{thm:path_bound}
For a path $\rho_0\to\rho_1$ over which $\rob(\rho_x)$ is an absolutely continuous function of $x$ and $W_x$ is any optimal robustness witness for $\rho_x$, the integrated variance obeys 
\begin{align}
\label{eq:BoundVariance}
    \int_{0}^{1} \Var(W_x) dx \geq \frac{(\Delta\rob)^2}{ \int_{0}^{1} I_x dx },
\end{align}
for quantum Fisher information $I_x = \Var(\logder_x)$ and total change in robustness $\Delta\rob = \rob(\rho_1)-\rob(\rho_0)$.
\end{theorem}
\end{comment}

\begin{figure}[!htbp]
    \centering
    \includegraphics[width = 0.42\textwidth]{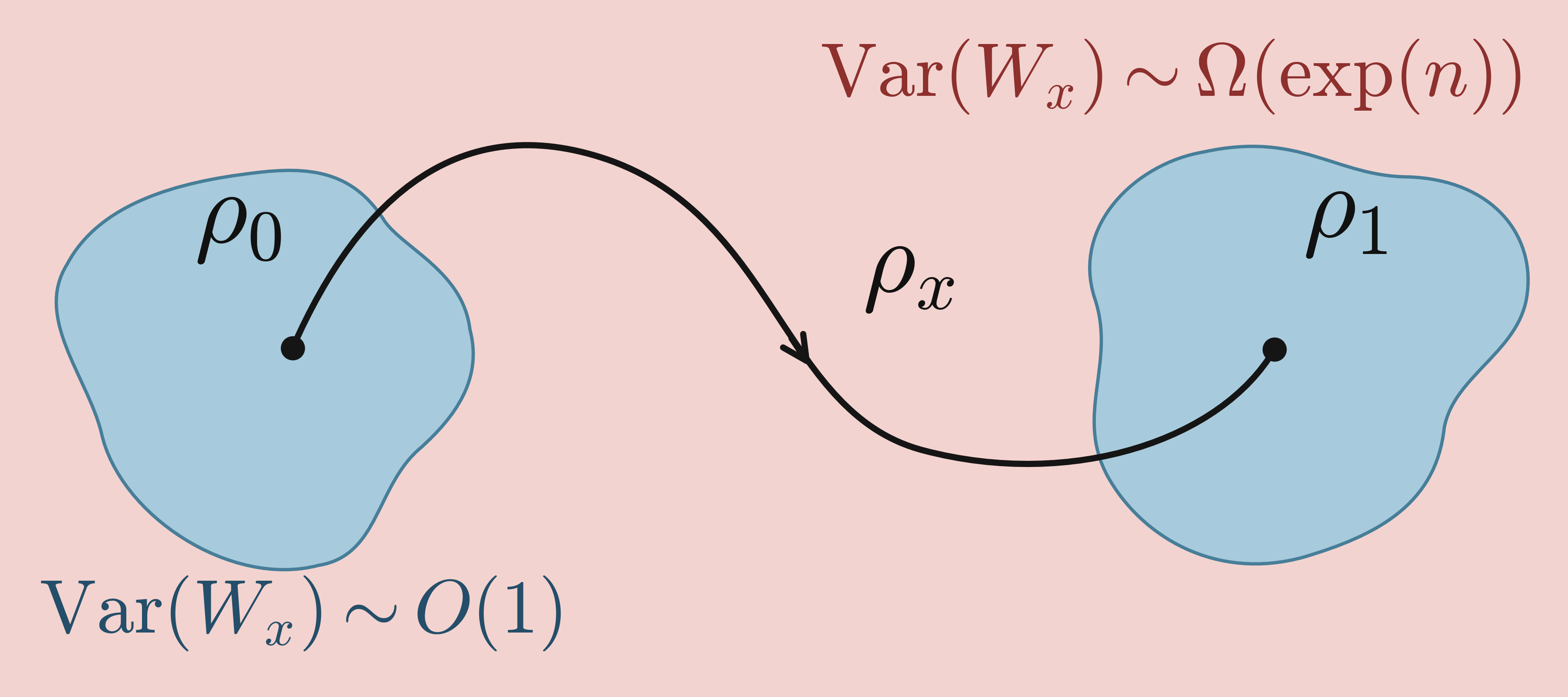}
    \caption{Consider any free state $\rho_0\in\free$ and exponentially-resourceful state $\rho_1$ with $\rob(\rho_1)\sim\exp(n)$. Any absolutely continuous path $\{\rho_x\ \vert\ 0\leq x\leq1\}$ of polynomial Fisher length which connects $\rho_0$ to $\rho_1$, must pass through a region within which every optimal witness has variance exponential in system size (red). This holds even if path endpoints admit low-variance optimal witnesses (blue regions). Thus, robustness cannot be efficiently measured on a subset of the path.}
    \label{fig:fig1}
\end{figure}

Consider probing the resource generation, in an experimentally measurable manner, of an $n$-qubit system evolving from $\rho_{0}$ to $\rho_{1}$ with $\Delta\rob\in\exp(n)$. When $I_x\in O(\mathrm{poly}(n))$ for $x$ along the path, we are guaranteed the integrated variance in \cref{eq:BoundVariance} is $\Omega(\exp(n))$. The latter condition occurs, for example, any time the generator $\logder_x$ of changes in $x$ is a sum of geometrically local terms over $k \in O(1)$ qubits, in which case $I_x \in O(n)$ for states with finite correlation length, or $O(n^2)$ for globally correlated states. Another case with polynomial Fisher information is real or imaginary time evolution generated by a local Hamiltonian in time $T\in\mathrm{poly}(n)$, for which $I_x\leq 4T^2\Var(H_x)\in O(\mathrm{poly}(n))$, where the factor of $T^2$ comes from the rescaling $x = t/T$~\cite{liu2019quantum}.
For some non-measure-zero subset of the path from $\rho_0$ to $\rho_1$, any estimator has exponential variance, and so robustness cannot be measured efficiently on this subset. This holds even when both endpoints possess an easily measurable, low-variance witness, in which case the high-variance subset separates the two states while not including them. Figure~\ref{fig:fig1} illustrates such a no-go result for efficiently and continuously tracking the creation of a quantum resource by robustness measures.

%Trying to update since ChatGPT finally explained why this isn't valid properly
%\brian{Work in progress; previous version had problem.}
\begin{comment}
V2: Moreover, any two regions $A,B$ of state space which both have exponentially large robustness but different exponential growth rates will have an exponentially large robustness gap. Then every path of sub-exponential Fisher length $\int_0^1 I_x dx$ from a state in $A$ to one in $B$ will have a subset on which the robustness cannot be efficiently estimated. This implies state space can be partitioned into regions of differing exponential robustness growth rate, between which one cannot efficiently track the growth of robustness along paths of sub-exponential Fisher length. 

More formally, let $A_n, B_n$ be $n$-dependent regions such that there exist polynomials $p_1,p_2$ for which $\forall \rho_{A}\in A_n,\ \rob(\rho_{A_n})\in[p_1(n)e^{b_A n},p_2(n) e^{b_A n}]$, ditto $B$, and assume $b_B \gneq b_A$. Then $\forall n$, $\forall \rho_A\in A_n, \rho_B\in B_n$, we have $\Delta\rob\in\Omega(p_1(n)e^{b_B n})$. When $\int_0^1 I_x dx \in O(\mathrm{poly}(n))$, \cref{eq:boundchange} guarantees that $\int_0^1 \Var(W_x) dx \in \Omega( e^{b'n})$ for some $b' > 0$. Since the path is of length $1$, this then implies $\Var(W_x) \Omega (e^{b' n})$ for some non-measure-zero subset of the path.
\end{comment}

Consider two regions $A$ and $B$ of state space between which there is an exponentially large robustness gap, i.e., $\min_{\rho_A\in A,\rho_B\in B}\abs{\rob(\rho_A)-\rob(\rho_B)}\in\exp(n)$. Then every path of sub-exponential Fisher length $\int_0^1 I_x dx$ from a state in $A$ to one in $B$ must cross through the set of states for which the robustness cannot be efficiently estimated. Moreover, any two regions $A,B$ with $\rob(\rho_A)\sim\mathrm{poly}(n)e^{b_A n}\ \forall \rho_A\in A$ and $\rob(\rho_B)\sim\mathrm{poly}(n)e^{b_B n} \forall \rho_B \in B$ for $b_A \neq b_B$ will have $\Delta\rob\sim \exp(n)$. This implies state space can be partitioned into regions of differing exponential robustness growth rate, between which any path either has exponential Fisher length or contains states for which robustness cannot be efficiently estimated.

If we instead consider the relative variance, the picture is much more positive. A slight modification of the previous bound bounds the relative variance in terms of ratio $\rob(\rho_1)/\rob(\rho_0)$:
\begin{align}
\label{eq:boundlogchange}
    \begin{split}
        \abs{\log\left(\frac{\rob(\rho_1)}{\rob(\rho_0)}\right)} &\leq \int_0^1\abs{\partial_x\log\rob(\rho_x)}dx\\
        &= \int_0^1 \frac{1}{\rob(\rho_x)} \abs{\partial_x\rob(\rho_x)}dx\\
        &\leq \int_0^1 \frac{1}{\rob(\rho_x)}\sqrt{\Var(\logder_x)\Var(W_x)} dx\\
        &\leq \sqrt{ \int_{0}^{1} I_x dx } \, \sqrt{ \int_{0}^{1} \frac{\Var(W_x)}{\rob(\rho_x)^2} dx },
    \end{split}
\end{align}
where we get from the second to third line by taking similar steps as in \cref{eq:boundchange}. Then the resulting bound on the relative variance is
\begin{align}
\label{eq:BoundLogVariance}
     \int_{0}^{1} \frac{\Var(W_x)}{\rob(\rho_x)^2} dx \geq \frac{ \abs{\log\frac{\rob(\rho_1)}{\rob(\rho_0)}}^2}{\int_{0}^{1} I_x dx}.
\end{align}

Under our convention $\rob \geq 1$, and $\rob \in O(\exp(n))$ for many resources of interest, such as magic, entanglement, generalized robustness of coherence, and for well-behaved resources on product states. In such cases, as long as $I_x \in \Omega(1)$ (non-decreasing in system size), the lower bound in \cref{eq:BoundLogVariance} goes as $O(n^2)$ allowing the possibility of estimating robustness to low relative error in $O(n^2)$ measurements.

To compare the two lower bounds, consider the generalized robustness of magic $\rob^+$ along a path of states of the form $\ket{\psi_x T^{n-1}}$, with $\ket{\psi_0}=\ket{+}$ and $\ket{\psi_1}=\ket{T}$, where $\ket{T}$ is defined as
\begin{align}
    \ket{T} \coloneqq T \ket{+} = \left(\ket{0} + e^{i\pi/4}\ket{1} \right)/\sqrt{2}.
\end{align}
Recall that for magic, $\ket{+}$ is a stabilizer (free) state, while $\ket{T}$ is resourceful. $\ket{T}$ can be generated by applying the single-qubit Hamiltonian $H = -\frac{\pi}{4}\ketbra{1}{1}$ to $\ket{+}$ for time $t=1$, and so $\int_0^1 I_x dx \in O(1)$. $\ket{+}$, $\ket{T}$, and in fact all states along the path between them, share a common optimal generalized robustness witness $W_T = \ketbra{T}{T}/F(T)$, proportional to a projector onto the $\ket{T}$ state~\cite{seddon2021quantifying}. The generalized robustness for the $\ket{T}$ state can then be measured with zero variance, since the state is being measured in its own eigenbasis.

Generalized robustness of magic is multiplicative for product single-qubit pure states, so in this case~\cite{seddon2021quantifying}
\begin{align}
\begin{split}
    \rob^+(\rho_0) &= \rob^+(T)^{n-1}\rob^+(+) = \rob^+(T)^{n-1}\\
    \rob^+(\rho_1) &= \rob^+(T)^{n}\\
    \Delta\rob^+ &= \rob^+(T)^{n-1}(\rob^+(T)-1)\in\Theta(\exp(n))\\
    \log\left(\frac{\rob^+(\rho_1)}{\rob^+(\rho_0)}\right) &= \log\rob^+(T)\in O(1).
\end{split}
\end{align}
This also implies the optimal witness is a product of the single state optimal witnesses, in this case $W_T^{\otimes n} = \ketbra{T^n}{T^n}/F(T)^n$, and $\rob^+(T)^n = 1/F(T)^n$.

The optimal witness variance at each point $x$ is 
\begin{align}
\begin{split}
    \Var_x(W_T^{\otimes n}) &= \frac{1}{F(T)^{2n}}\left(\abs{\braket{T\vert\psi_x}}^2-\abs{\braket{T\vert\psi_x}}^4\right)\\
    &= \rob^+(T)^{2n-2}\Var_{\psi_x}(W_T).
\end{split}
\end{align}

The scaling captured by both bounds is tight, as $\Var_x(W_T^{\otimes n}) \sim \exp(n)$ while $\Var_x(W_T^{\otimes n})/\rob^+(\rho_x)^2 = \Var_{\psi_x}(T)/\rob^+(\psi_x)^2\in O(1)$ at each point. However, if we interpret the $n$-qubit system as $n$ independent qubits, we see that the path can be interpreted as creating only one qubit's worth of resource, so the exponential scaling in $\Delta\rob^+$ is somewhat misleading. Similarly, witness measurements on $\rho_x$ can be interpreted as $n-1$ independent zero-variance measurements and one with nonzero variance (for $x<1$), so the exponential scaling of absolute variance is somewhat misleading, since it comes from the scale of $W_T^{\otimes n}$ alone. The relative variance does not have such shortcomings, highlighting why estimating $\rob^+$ to low multiplicative error can be qualitatively easier than to low absolute error.

\section{Discussion}
\label{sec:Discussion}

We obtain bounds on the sampling cost of estimating standard and generalized robustness that apply across resource theories. These bounds expose statistical limitations that remain even when an optimal witness is known, and connect experimental precision to the target state's resource content. Our results indicate potential hurdles to resource estimation on mixed states, estimation of standard robustness, and resource estimation to low absolute error. This suggests that the estimation of generalized robustness to low relative error on highly resourceful pure states is a preferable candidate resource-estimation task, since it avoids many of the hurdles.

Treating the optimal witness as an observable allows one to place Cram\'er-Rao bounds on the number of measurements required to estimate robustness for a range of resources (Theorems \ref{thm:CR_mixed}, \ref{thm:CR_pure}). For mixed states, this shot complexity generically scales at least quadratically in the resource robustness, making resource estimation of generic highly resourceful states inefficient. For generalized robustness of pure states, the shot complexity lower bound obeys different scaling -- linear in robustness for state families with fixed robustness-fidelity ratio. When the task is estimating the generalized robustness to fixed relative error, the bound has a decreasing relationship with resource robustness, indicating the task has the potential to be easy for highly resourceful states. 

Our geometric bounds highlight that shot complexity can remain large even for low-resource states, depending on the structure of the free set. In particular, the bound in \cref{thm:var_bound_r-R} is tight when active free states are orthogonal, which is the case for the generalized robustness of coherence and bipartite entanglement. In these cases, one can analytically find a witness which both is optimal and achieves minimal shot complexity. A stronger obstruction arises when tracking resource generation along a trajectory in state space (\cref{thm:path_bound}). A large gain in robustness necessitates a path which crosses through intermediate states with high witness variance, even when the path endpoints are easily measurable. This establishes a no-go theorem in tracking the onset of resource robustness in state-space evolutions with fixed absolute error. The corresponding bound on relative error is significantly weaker, scaling squared-logarithmically in resource gain. As we saw from the example, high absolute error and low relative error can coexist for local evolutions on highly resourceful states. This distinction highlights the importance of the specific estimation task when assessing the feasibility of a benchmark. 

An open question is to what extent the Cram\'er-Rao bounds in \cref{sec:CR-bounds} can be improved without requiring prior knowledge about the optimal witness. A perturbation in the direction of an optimal witness yields a saturable Cram\'er-Rao bound, but the resulting expression is witness-dependent, limiting its practicality. A useful next step is to find perturbations determined by accessible state properties which yield tighter bounds. The geometric bounds suggest a complementary approach: restricting the allowed set of potential optimal witnesses for a given state would let one better bound its distance from its optimal witness, resulting in tighter geometric bounds on shot complexity. 

The location and extent of the unmeasurable subset observed along resource-generating paths described in \cref{sec:Sampling} also remain unclear. Relative variance bounds and the absolute variance upper bound imply typical unmeasurable subsets consist of highly-but-not-maximally resourceful states. 
%However, it remains open whether they make up a thin boundary region which paths briefly cross, or if generic states cannot have their resource robustness measured efficiently. 
However, it remains open whether the unmeasurable subset is inherently path-dependent, whether there exists a boundary region of unmeasurable states separating measurable regions, or if efficiently measurable states make up small islands in state space amid a sea of unmeasurable states. 
This question motivates an experimental approach where one evolves a state from zero to exponentially high resource, while measuring resource at each step to track when the witness variance becomes prohibitively high. Generalized robustness of coherence is a particularly clean choice for this, as the optimal witness can be found analytically, and potentially in practice for all states along the trajectory. Comparisons of shot complexity and computational complexity of finding the optimal witness would also be revealing -- categorizing the set of states which are low in both complexity metrics could guide experiments towards efficient demonstration of quantum capabilities.

\section*{Acknowledgments}

This work was supported by the Quantum Science Center (QSC), a National Quantum Information Science Research Center of the U.S. Department of Energy (DOE). 
B.B. acknowledges support from the DOE, Office of Science, Accelerated Research in Quantum Computing, Fundamental Algorithmic Research toward Quantum Utility (FAR-Qu).
L.P.G.P. acknowledges support from the DOE, Office of Science, Basic Energy Sciences program (award No. DE-SCL0000157).

ChatGPT-6 Astra was used for editing, literature search, and to ensure the accuracy and novelty of results. 
It assisted in the derivations of \cref{eq:witness-purification}, \cref{lem:witness_fidelity}, and \cref{thm:var_bound_inst}.
%It generalized results by deriving \cref{eq:OptimalWitnessPure} and extending \cref{lem:witness_fidelity} from the $r=2$ lower bound, and helped find the non-trigonometric form of \cref{thm:var_bound_inst}. 
It derived the results in Appendix \ref{app:scope}, helped shorten Appendix \ref{app:CR-proof-GR}, and made \cref{fig:fig1}.

\bibliography{magic, bib}

\widetext

%\clearpage
%\end{document}

\section*{APPENDIX}

\appendix
\begin{appendix}
\numberwithin{equation}{section}

\section{Resource Cram\'er-Rao Bounds}
\label{app:CR-proofs}
This section contains derivations of the perturbative Cram\'er-Rao (CR) bounds on generalized robustness and on moment functions used in \cref{sec:CR-bounds}.

\subsection{Generalized Robustness of Pure States}
\label{app:CR-proof-GR}

%ChatGPT's chortening + extension to mixed free states miximizing Fidelity
Here we derive a perturbative CR bound on the variance of any optimal witness of generalized robustness of pure states. For pure state $\ket{\psi}$, consider the perturbed state $\ket{\psi_c} = \ket{\psi} + c\ket{\psi'}$ in the limit $c\rightarrow0$, where $\braket{\psi\vert\psi'}=0$ and $c \in \mathbb{R}$. We use the perturbation $\ket{\psi'} = (F - \phi)\ket{\psi}$, where $F$ is the free-state fidelity of $\ket{\psi}$ and $\phi$ is the (potentially mixed) free state maximizing $\braket{\psi\vert\phi\vert\psi}$, so that $F = \braket{\psi\vert\phi\vert\psi}$. At $c=0$, the symmetric log-derivative is $\logder_c\big\vert_{c=0} = 2(F\mathbb{I}-\phi)$, making the quantum Fisher information
\begin{align}
\begin{split}
    I_c \big\vert_{c=0} &= \braket{\logder_c^2}\big\vert_{c=0} = 4(\braket{\phi^2}-F^2)\leq 4F(1-F).
\end{split}
\end{align}

Turning now to the numerator of the CR bound, we have 
\begin{align}
\partial_c\rob^+_W(\psi_c)\big\vert_{c=0} = 2\Re{\braket{\psi\vert W\vert\psi}} = 2(\rob^+ F - \Re{\braket{\psi\vert W\phi\vert\psi}})    
\end{align}

Cauchy-Schwarz gives $\abs{\braket{\psi\vert W\phi\vert\psi}}^2 \leq \braket{\psi\vert W\vert\psi}\braket{\psi\vert\phi W\phi\vert\psi} = \rob^+\tr{W \phi \psi \phi}$ for $\psi:=\ketbra{\psi}{\psi}$, while trace-norm submultiplicativity gives
\begin{align}
    \tr{W \phi \psi \phi}\leq \tr{\abs{W \phi \psi \phi}} \leq \tr{\abs{\sqrt{\phi}W\sqrt{\phi}}}\tr{\abs{\sqrt{\phi}\psi\sqrt{\phi}}}=\tr{W\phi}\tr{\psi\phi}\leq F.
\end{align}
Thus, $\braket{\psi\vert\phi W\phi\vert\psi} \leq \sqrt{\rob^+ F}$. We can then bound 
\begin{align}
\begin{split}
    \abs{\partial_c\rob^+_W(\psi_c)}^2\big\vert_{c=0} &\geq 4(\rob^+ F - \Re{\braket{\psi\vert W\phi\vert\psi}})^2\geq 4(\rob^+ F - \abs{\braket{\psi\vert W\phi\vert\psi}})^2\geq 4(\rob^+F - \sqrt{\rob^+F})^2\\
    &= 4\rob^+ F(\sqrt{\rob^+ F}-1)^2,
    \end{split}
\end{align}
since $\rob^+ F \geq 1$ implies $\rob^+ F - \Re{\braket{\psi\vert W\phi\vert\psi}} \geq \rob^+ F - \sqrt{\rob^+ F} \geq 0$. Plugging this and the previous Fisher information result into the CR bound in \cref{eq:BG-CR} yields
\begin{align}
    \Var(W) \geq \frac{4\rob^+ F(\sqrt{\rob^+ F} -1)^2}{4 F (1-F)},
\end{align}
which after canceling terms gives the final form in \cref{thm:CR_pure}. While this is not the tightest CR bound which can be placed on $\Var(W)$ (which would come from optimizing over path-derivatives), it is the one we found which can be expressed in terms of standard, index-free quantities of $\ket{\psi}$.

\subsection{Moments}
\label{app:CR-moments}

A number of resource measures can be described as the $k^{th}$ moment of some distribution, including purity, entanglement purity, and stabilizer moments. Moments are interesting in their own right, but are also useful as building blocks of entropy measures, such as stabilizer R\'enyi entropy~\cite{tirrito2024quantifying, leone2022stabilizer}. While they are not linear in state like robustness is, they can be written as the expectation value of a joint observable acting on $k$ state copies~\cite{ekert2002direct}. For standard purity $P_k$ (entanglement purity $E_k$), the relevant measurement operator is the SWAP (partial SWAP) gate acting on $k=2$ copies, while the $k^{th}$ stabilizer moment $A_k$ is instead written in terms of $\Gamma = 2^{-n}\sum_{\vec p} \sigma_{\vec p}^{\otimes 2k}$ acting on $2k$ state copies~\cite{haug2024efficient, leone2022stabilizer}. As these measurements do not necessarily estimate the quantity with minimal error, we turn to the CR bound once more to investigate if better measurements on $k$ state copies exist.

To approach this, we first define a formalized moment function $F_k$, acting on state $\rho$ and a (potentially $\rho$-dependent) set of observables $\{M_j\}_j$ as
\begin{align}
    F_k(\{M_j\}_j) = \sum_j \braket{M_j}^k.
\end{align}
The previously mentioned examples are special cases of this function: the purity moment $P_k = \tr{\rho^k} = F_k(\{\ketbra{\psi_j}{\psi_j}\})$, where $\{\ket{\psi_j}\}$ are eigenvectors of $\rho$, the entanglement purity moment $E_k = \sum_j \lambda_j^{k} = F_k(\{\ketbra{u_j}{u_j}\otimes\mathbb{I}\}_j)$, for Schmidt decomposition $\ket{\psi} = \sum_j \sqrt{\lambda_j}\ket{u_j v_j}$ when $\rho$ is pure, and the stabilizer moment $A_k = 2^{-n}\sum_p \braket{\sigma_p}^{2k} = 2^{-n}F_{2k}(\{\sigma_p\})$, for $n$-qubit Pauli strings $\sigma_p$. Thus by bounding the variance of an estimator of $F_k$, we can determine the difficulty of estimating these functions.

Unlike in previous cases, we start from the symmetric log-derivative. Consider a state $\rho_c$ perturbed by the log-derivative $\logder_c = 2(G_k - F_k)$, so that $\partial_c\rho_c = \{G_k,\rho_c\}-2F_k\rho_c$ and $\Var(\logder_c) = 4\Var(G_k)$. For a given set of observables $\{M_j\}$, which we will hold fixed, define $G_k = \sum_j \braket{M_j}^{k-1} M_j$, so that $F_k = \braket{G_k}$. Then
\begin{align}
\begin{split}
    \partial_c F_k (\{M_j\}_j)\Big\vert_{c=0} &= k\sum_j\braket{M_j}_{\rho}^{k-1}\tr{M_j \partial_c\rho_c}\Big\vert_{c=0} = k\tr{G_k\partial_c\rho_c}\Big\vert_{c=0} = k\tr{G_k\{G_k,\rho\}-2G_k\rho F_k}\\
    &= 2k(\braket{G_k^2} - F_k^2) = 2k\Var(G_k).
\end{split}
\end{align}
When considering $k$ state copies $\rho_c^{\otimes k}$, it follows from the product rule that the full log-derivative is $\logder_{c,k} = \sum_i \logder_{c,S_i}$, the sum of log-derivatives acting on each system copy $S_i$. %, since
%\begin{align}
%    \begin{split}
%        \frac{1}{2}\{\logder_{c,k},\rho^{\otimes k}_c\} &= \partial_c (\rho_c^{\otimes k}) = \sum_i (\partial_c \rho_{c, S_i})\bigotimes_{j\neq i}\rho_{S_j} = \frac{1}{2}\sum_i \{\logder_{c,S_i},\rho_c^{\otimes k}\}.
%    \end{split}
%\end{align}
Given that $\braket{\logder_{c,S_i}\logder_{c,S_j}} = \braket{\logder_c^2}\delta_{ij}$, we find that
\begin{align}
    \Var(\logder_{c,k}) = \sum_{ij}\braket{\logder_{c,S_i}\logder_{c,S_j}} = k\Var(\logder_c) = 4k\Var(G_k).
\end{align}
Then for some operator $M$ acting on $k$ state copies with $F_k = \tr{M \rho^{\otimes k}}$, CR tells us that
\begin{align}
    \begin{split}
        \Var(M) &\geq \frac{\abs{\partial_c F_k}^2}{\Var(\logder_{c,k})} = \frac{4 k^2 \Var(G_k)^2}{4 k \Var(G_k)} = k\Var(G_k).
    \end{split}
\end{align}

While we can write $\Var(G_k) = \braket{G_k^2}-F_k^2 = \sum_{ij}\braket{M_i}^{k-1}\braket{M_j}^{k-1}\Cov(M_i,M_j)$, it is hard to cleanly solve for the $\braket{G_k^2}$ term in general. When $0 \leq M_i \leq \mathbb{I}$ we can upper bound $\abs{\braket{M_i M_j}}\leq 1$ and know $\abs{\braket{M_i}}=\braket{M_i}$, so $\braket{G_k^2}\leq F_{k-1}^2$ by Cauchy-Schwarz. Moreover, when $\{M_i\}$ are orthogonal operators so that $\tr{M_i M_j} = \delta_{ij}$, we can bound $\braket{G_k^2} \leq \tr{G_k^2} = F_{2k-2}$. This is the relevant case for $A_k$, up to a factor of $2^n$ since $\tr{\sigma_p\sigma_{p'}}=2^n\delta_{p p'}$. In particular, when $\{M_i\}$ form a set of orthogonal projectors (as they do for $P_k,\ E_k$), $\braket{M_i M_j} = \braket{M_i}\delta_{ij}$, $\braket{G_k^2} = F_{2k-1}$ exactly, and $\Var(G_k) = F_{2k-1}-F_k^2$. When $\ket{\psi}$ is an eigenvector of $G_k$ with maximal eigenvalue, we have $G_k\ket{\psi}=F_k\ket{\psi}$. In this case $\Var(G_k) = \braket{G_k^2} - F_k^2 = 0$, so $F_k$ can be measured with zero variance by measuring $G_k$.

\section{Bound Tightness Examples}
\label{app:scope}

Several families of states exist for which the optimal witness can be found explicitly, allowing us to test the tightness of the bounds in Sections \ref{sec:CR-bounds} and \ref{sec:geometry}. 

\subsection{Mixed State Robustness}
\label{app:scope_mixed}
While the bound in \cref{thm:CR_mixed} approaches triviality for nearly pure states, it is tight in various cases, such as for robustness of entanglement of Werner states, dephased maximally entangled states, and isotropic states. We will focus on the third example here. Isotropic states are defined as \cite{jafarizadeh2005exact}:
\begin{align}
    \rho_f = f\mathrm{P} + \frac{1-f}{d^2-1}\left(\mathbb{I}-\mathrm{P}\right)\hspace{.5in}\mathrm{P} = \ketbra{\Phi}{\Phi},
\end{align}
for maximally entangled state $\ket{\Phi} = \frac{1}{\sqrt{d}}\sum_i\ket{ii}$, $d$ the dimension of a single subsystem, and $\frac{1}{d} < f < 1$. With our normalization, the robustness of bipartite entanglement is $\rob(\rho_f) = 2df-1$~\cite{jafarizadeh2005exact}. Any separable (i.e., free) state $\phi$ can be decomposed as $\phi = \sum_i p_i \ketbra{a_i b_i}{a_i b_i}$ for $\sum_i p_i = 1$, so one has
\begin{align}
\label{eq:separable-cauchy}
    \tr{\mathrm{P}\phi} &= \sum_i \frac{p_i}{d} \abs{\sum_j\braket{jj\vert a_i b_i}}^2 \leq \sum_i \frac{p_i}{d} \sum_j\abs{\braket{j\vert a_i}}^2\sum_k\abs{\braket{k\vert b_i}}^2 = \frac{1}{d}\sum_i p_i = \frac{1}{d},
\end{align}
where the inequality comes from Cauchy-Schwarz, and the following equality from the fact that $\sum_i\ketbra{i}{i} = \mathbb{I}$. Thus, the witness $W^*=2d\mathrm{P}-\mathbb{I}$ is feasible, as $\abs{\tr{W^*\phi}} \leq 1 $ for separable $\phi$. Moreover, one can verify that $\tr{W^*\rho_f} = 2df-1$, so $W^*$ is optimal. One can then explicitly calculate $\Var(W^*) = 4d^2f(1-f)$. Turning to the bound in \cref{thm:CR_mixed}, we can use the fact that $\mathrm{P},\ \mathbb{I}-\mathrm{P}$ are orthogonal projectors and $\rho_f$ is full rank to calculate $\rho_f^{-1}$, and pick the optimal separable state $\phi = \rho_{\frac{1}{d}}$. Substituting this in and simplifying reveals that the bound in~\cref{thm:CR_mixed} is equal to $\Var(W^*)$. Thus, for this class of states the bound is tight.

\subsection{Pure-state Generalized Robustness}
\label{app:scope_pure}
We consider the case of generalized robustness of coherence for simplicity. Equivalently, the results hold for generalized robustness of bipartite entanglement, which has similar properties since the Schmidt decomposition acts as an orthonormal basis (as opposed to standard robustness of coherence and entanglement, which do not behave similarly). 

%Claude wants me to comment that robustness of coherence is infinite. I prefer this as-is

As previously mentioned, \cref{thm:var_bound_r-R} yields a tight lower bound in this case, since a target state $\ket{\psi} = \sum_{i=1}^r c_i \ket{\phi_i}$ for $\ket{\phi_i}\in\free$ admits the variance-minimizing optimal witness $W^* = \ketbra{v}{v}$, with $\ket{v} = \sum_{i=1}^r \ket{\phi_i}$. This witness has variance $\Var(W^*) = \rob^+(r-\rob^+)$, equal to the bound since $\theta_f = \frac{\pi}{2}$ and $F_r = \frac{1}{r}$ for coherence and bipartite entanglement. The other generalized robustness bounds in this paper are not always tight for coherence, but are worth comparing since they perform better in other regimes, such as for magic where the bound in \cref{thm:var_bound_r-R} becomes trivial whenever $\rob^+ > \sqrt{2}$.

\subsubsection{Low-resource state}
Let $\ket{\psi^-_\epsilon}$ be a nearly free state, defined as $\ket{\psi^-_\epsilon} = \sqrt{1-\epsilon^2}\ket{\phi_1} + \epsilon\sum_{i=2}^r c_i\ket{\phi_i}$, where we take $c_i > 0$ and $\sum_{i=2}^r c_i^2 = 1$ without loss of generality. Note that orthogonality of $\{\ket{\phi_i}\}_i$ means the decomposition is unique, and thus optimal for pure-state extent $\xi$ [\cref{eq:purestate-extent}]. Let $A = \sum_{i=2}^r c_i \leq \sqrt{r-1}$, with equality when all $c_i = 1/\sqrt{r-1}$. For states $\ket{\psi_\epsilon^-}$, $\rob^+ = \xi = (\sqrt{1-\epsilon^2}+A\epsilon)^2$ and $F = 1-\epsilon^2$. Let $\mathrm{CR}_2$, $\mathrm{Geo}_1$, and $\mathrm{Geo}_2$ denote the variance bounds in Theorems \ref{thm:CR_pure}, \ref{thm:var_bound_r-R}, and \ref{thm:var_bound_inst}, respectively, and $V_\mathrm{min}$ the minimum achievable variance by an optimal witness. Plugging in the resolution of $\rob^+, F, F_r$ and expanding to first order in $\epsilon$ yields
\begin{align}
    \begin{split}
        V_\mathrm{min} = \mathrm{Geo}_1 &= (r-1) + 2A(r-2)\epsilon + O(\epsilon^2), \\
        \mathrm{Geo}_2 &= (r-1) + 2\sqrt{r-1}(2A\sqrt{r-1}-r)\epsilon + O(\epsilon^2), \\
        \mathrm{CR}_2 &= A^2 + 2A(A^2-1)\epsilon + O(\epsilon^2).
    \end{split}
\end{align}
$\mathrm{Geo}_2$ is tight in the limit $\epsilon \to 0$, and differs by of $V_\mathrm{min} - \mathrm{Geo}_2 = 2r(\sqrt{r-1}-A)\epsilon+O(\epsilon^2)$ otherwise. Meanwhile, $\mathrm{CR}_2$ has an $O(1)$ difference in general, and has an optimality ratio of $\mathrm{CR}_2 / V_\mathrm{min} = A^2/(r-1)+O(\epsilon)$. This ratio goes to one when $c_i = 1/\sqrt{r-1}\ \  \forall i$, in which case $\mathrm{CR}_2$ is tight for all $\epsilon$. Intuitively, the $O(1)$ difference in $\mathrm{CR}_2$ comes from the fact that we took a CR bound in the direction $\ket{\phi}-\ket{\psi^-_\epsilon}$ rather than the optimal $\ket{\omega}-\ket{\psi^-_\epsilon}$ for optimal witness state $\ket{\omega}$. Taking the latter in fact yields \cref{eq:var_bound_derivation_geometric}, which is tight but $\ket{\omega}$-dependent.

\subsubsection{High-resource state}
Let $\ket{\psi^+_\epsilon}$ be a nearly-maximally-resourceful state, defined (for coherence, or bipartite entanglement via the Schmidt decomposition) as $\ket{\psi^+_\epsilon} = \sqrt{1-\epsilon^2}\ket{+_r} + \epsilon\ket{u}$, where $\ket{+_r} = \frac{1}{\sqrt{r}}\sum_{i=1}^r\ket{\phi_i}$ and $\ket{u}$ is some normalized vector obeying $\braket{+_r\vert u}=0$. Let $u_\mathrm{max} = \max_i{\Re[u_i]}$ be the maximal real coefficient of $\ket{u}$ in the free state basis, and $u_\mathrm{sum} = \sum_{i=1}^r \Re[u_i]^2$ for convenience. In this case we have $\rob^+ = \xi = r(1-u_\mathrm{sum} \epsilon^2)+O(\epsilon^3)$ and $F = \frac{1}{r}+\frac{2 u_\mathrm{max}}{\sqrt{r}}\epsilon+O(\epsilon^2)$. Plugging these into the variance bounds gives
\begin{align}
    \begin{split}
        V_\mathrm{min} &= \mathrm{Geo}_1 = r^2 u_\mathrm{sum}\epsilon^2+O(\epsilon^3), \\
        \mathrm{Geo}_2 &= \frac{r^3 u_\mathrm{max}^2}{r-1}\epsilon^2 + O(\epsilon^3), \\
        \mathrm{CR}_2 &= \frac{r^3 u_\mathrm{max}^2}{r-1}\epsilon^2 + O(\epsilon^3),
    \end{split}
\end{align}
where $\mathrm{CR}_2$ and $\mathrm{Geo}_2$ differ by $O(\epsilon^3)$ terms. Evidently, both have an optimality ratio of $\frac{r u_\mathrm{max}^2}{(r-1)u_\mathrm{sum}}+O(\epsilon)$, which achieves one when $u_\mathrm{max}$ takes its largest possible value, $u_\mathrm{max} = \sqrt{\frac{r-1}{r}}$. 

\subsubsection{State family with tight bounds}
In both cases, the limit where $\mathrm{CR}_2 = \mathrm{Geo}_2 = V_\mathrm{min}$ occurs when $\ket{\psi^-_\epsilon}$ and $\ket{\psi^+_\epsilon}$ approach a family of states for which all bounds considered here are always equal and optimal. That family of states is given by $\ket{\psi_f} = f\ket{\phi_1}+\sqrt{1-f^2}\ket{+_{2,r}}$ for $\ket{+_{2,r}} := \frac{1}{\sqrt{r-1}}\sum_{i=2}^r\ket{\phi_i}$ and $f \geq 1/\sqrt{r}$. In this case $\rob^+ = (f + \sqrt{(r-1)(1-f^2)})^2$ and $F = f^2$. Calculating all bounds eventually yields the same result for this family, though the result itself is unenlightening.

\end{appendix}
\end{document}